\documentclass[prd,aps,nofootinbib,floatfix,onecolumn,10pt]{revtex4}
\usepackage{multirow}
\usepackage{textcomp}
\usepackage{amsmath,graphicx,color,epsfig}

\begin{document}
\title{Flavor SU(3) Topological Diagram and Irreducible Representation Amplitudes for Heavy Meson Charmless Hadronic Decays: Mismatch and Equivalence}
\author{Xiao-Gang He$^{1,2,3}$~\footnote{Email:hexg@sjtu.edu.cn}, and  Wei Wang $^{1}$~\footnote{Email:wei.wang@sjtu.edu.cn}}
\affiliation{
 $^1$ T.-D. Lee Institute  and INPAC,  Shanghai Key Laboratory for Particle Physics and Cosmology, MOE Key Laboratory for Particle Physics, School of Physics and Astronomy, \\ Shanghai Jiao Tong University, Shanghai  200240 \\
 $^2$ Department of Physics, National Taiwan University, Taipei 106\\
$^3$ National Center for Theoretical Sciences, TsingHua University,  Hsinchu 300}

\begin{abstract}
Flavor SU(3) analysis of heavy meson ($B$ and $D$) hadronic charmless decays can be formulated  in two {  different} ways. One is to construct the SU(3) irreducible representation amplitude (IRA) by decomposing effective Hamiltonian according to the SU(3) transformation  properties. The other is to use the topological diagrams (TDA). These two methods should give equivalent  physical results in the SU(3) limit. Using    $B \to PP$ decays as an example,  we point out  that previous analyses in the literature using these two methods  do not match consistently in several ways, in particular a few SU(3) independent amplitudes have been overlooked in the TDA approach.  Taking these new amplitudes into account, we find a consistent description in both schemes. These new  amplitudes can affect  direct CP asymmetries in some channels significantly. A consequence  is that for any charmless hadronic  decay of heavy meson, the direct CP symmetry cannot be identically zero.  
\end{abstract}
\maketitle

%

\section{Introduction} 

Hadronic charmless $B$ decays provide an ideal platform to extract the CKM matrix elements, test the standard model description of   CP violation  and look for new physics effects beyond the standard model (SM). Experimentally, quite a number of physical  observables like  branching fractions, CP asymmetries and polarizations have been precisely measured  by experiments at the electron-position colliders and hadron colliders. For a collection of these results, please see Refs.~\cite{Amhis:2016xyh,Patrignani:2016xqp}.    On the other hand, theoretical {  calculations of the decay amplitudes} greatly  rely on the factorization ansatz.  Depending on the explicit realizations of factorization, several QCD-based dynamic approaches have been established, such as QCDF~\cite{Beneke:2001ev}, PQCD~\cite{Keum:2000ph,Keum:2000wi,Lu:2000em}, SCET~\cite{Bauer:2000yr,Bauer:2001cu}. Apart from factorization approaches, the flavor SU(3) symmetry has been also wildly used in two-body and three-body heavy meson decays~\cite{Zeppenfeld:1980ex,Savage:1989ub,Deshpande:1994ii,He:1998rq,He:2000ys,Hsiao:2015iiu,Chau:1986du,Chau:1987tk,Chau:1990ay,Gronau:1994rj,Gronau:1995hm,Cheng:2014rfa}. 
An advantage of this method is its independence on the detailed dynamics in factorization.  Since the SU(3) invariant amplitudes can be determined by fitting the data,   the SU(3) analysis also provides a bridge between the experimental data and the dynamic approaches.

In the literature, the SU(3) analysis has been formulated in two    distinct ways. One is to derive the decay amplitudes correspond to various topological diagrams (TDA)~\cite{Chau:1986du,Chau:1987tk,Chau:1990ay,Gronau:1994rj,Gronau:1995hm,Cheng:2014rfa}, and another is to construct the SU(3) irreducible representation amplitude (IRA) by decomposing effective Hamiltonian according to irreducible representations~\cite{Savage:1989ub,Deshpande:1994ii,He:1998rq,He:2000ys,Hsiao:2015iiu}. These two methods should give the same physical results in the SU(3) limit when all relevant contributions are taken into account. However, as we will show   we find that previous analyses in the literature using these two methods  do not match consistently in several ways, in particular a few SU(3) independent amplitudes have been overlooked in the TDA approach for a heavy meson   decaying  into two light pseudoscalar SU(3) octet (or U(3) nonet) mesons.  In this work, we carry out a systematic analysis  and  identify possible missing amplitudes in order to establish the consistence between  the RRA and TDA approaches. We find that these new  amplitudes are sizable and may  affect  direct CP asymmetries in some channels significantly. An important  consequence of the inclusion of these amplitudes  is that for any charmless hadronic  decay  of heavy mesons, the direct CP symmetry cannot be identically zero, though  in some cases it is  tiny. 

The rest of this paper is organized as follows. In Sec.~\ref{sec:basics}, we introduce the SU(3) analysis using the TDA and IRA approaches. We  summarize those amplitudes already discussed in the literature. In Sec.~\ref{sec:solution}, we first point out the mismatch problem, and   then identify those missed amplitudes.  The complete sets of SU(3) independent amplitudes in both IRA and TDA approaches will be given  to establish equivalence of these two approaches. In Sec.~\ref{sec:conclusion}, we include the missing amplitudes to discuss the implications for hadronic charmless decays of $B$ and $D$ and draw our conclusions.

\section{  Basics for IRA and TDA approaches}
\label{sec:basics}

\subsection{SU(3) STRUCTURE} 

We start with the electroweak effective Lagrangian for hadronic charmless $B$ meson decays in the SM.
The  Hamiltonian ${\cal H}_{eff}$ responsible for such kind of decays at one loop level in electroweak interactions is given by~\cite{Buchalla:1995vs,Ciuchini:1993vr,Deshpande:1994pc}:
 \begin{eqnarray}
 {\cal H}_{eff}  &=& \frac{G_{F}}{\sqrt{2}}
     \bigg\{ V_{ub} V_{uq}^{*} \big[
     C_{1}  O_{1}
  +  C_{2}  O_{2}\Big]- V_{tb} V_{tq}^{*} \big[{\sum\limits_{i=3}^{10}} C_{i}  O_{i} \Big]\bigg\}+ \mbox{h.c.} ,
 \label{eq:hamiltonian}
\end{eqnarray}
where  $O_{i}$ is a four-quark operator or a moment type operator. The four-quark operators $O_i$ are given as follows:
\begin{eqnarray}
 O_1 = (\bar q^i u^j)_{V-A} (\bar u^j b^i)_{V-A}, && O_2 = (\bar q u)_{V-A} (\bar u b)_{V-A}, \nonumber\\
 O_3= (\bar q b)_{V-A} \sum_{q'} (\bar q'q')_{V-A}, && O_4= (\bar q^i b^j)_{V-A} \sum_{q'} (\bar q^{\prime j}q^{\prime i})_{V-A}, \nonumber\\
 O_5= (\bar q b)_{V-A} \sum_{q'} (\bar q'q')_{V+A}, && O_6= (\bar q^i b^j)_{V-A} \sum_{q'} (\bar q^{\prime j}q^{\prime i})_{V+A}, \nonumber\\
 O_7=\frac{3}{2} (\bar q b)_{V-A} \sum_{q'} e_{q'}(\bar q'q')_{V+A}, && O_8= \frac{3}{2}(\bar q^i b^j)_{V-A} \sum_{q'}e_{q'} (\bar q^{\prime j}q^{\prime i})_{V+A}, \nonumber\\
 O_9=\frac{3}{2} (\bar q b)_{V-A} \sum_{q'}e_{q'}  (\bar q'q')_{V-A}, &&  O_{10}= \frac{3}{2}(\bar q^i b^j)_{V-A} \sum_{q'}e_{q'}  (\bar q^{\prime j}q^{\prime i})_{V-A}. \label{operators}
\end{eqnarray}
In the above the $q$ denotes a $d$ quark for the $b\to d$ transition  or an $s$ quark for the $b\to s$ transition, while $q'=u,d,s$. 

At the hadron level, QCD penguin operators  behave as the ${\bf  \bar 3}$ representation while  tree and electroweak penguin  operators can
be decomposed in terms of a vector $H_{\bf \bar 3}$, a traceless
tensor antisymmetric in upper indices, $H_{\bf6}$, and a
traceless tensor symmetric in   upper indices,
$H_{\bf{\overline{15}}}$.
For the $\Delta S=0 (b\to d)$ decays, the non-zero components of the effective Hamiltonian are~\cite{Savage:1989ub,He:2000ys,Hsiao:2015iiu}:
\begin{eqnarray}
 (H_{\bf \bar3})^2=1,\;\;\;(H_{6})^{12}_1=-(H_{6})^{21}_1=(H_{6})^{23}_3=-(H_{6})^{32}_3=1,\nonumber\\
 2(H_{\overline{15}})^{12}_1= 2(H_{\overline{15}})^{21}_1=-3(H_{\overline{15}})^{22}_2=
 -6(H_{\overline{15}})^{23}_3=-6(H_{\overline{15}})^{32}_3=6,\label{eq:H3615_bb}
\end{eqnarray}
and  all other remaining entries are zero. For the $\Delta S=-1(b\to s)$
decays the nonzero entries in the $H_{\bf{\bar 3}}$, $H_{\bf 6}$,
$H_{\bf{\overline{15}}}$ can be  obtained from Eq.~\eqref{eq:H3615_bb}
with the exchange  $2\leftrightarrow 3$ corresponding to the $d \leftrightarrow s$ exchange. 

The above Hamiltonian can induce a $B_i$ meson to decay into two light pseudoscalar  nonet  $M^i_j$ mesons. There are three $B$ mesons 
$(B_i) = (B(\bar b u), B(\bar b d), B(\bar b s))$ which form a flavor SU(3) fundamental representation $3$. 
The light pseudoscalar mesons $M^i_j$ contain nine hadrons: 
\begin{eqnarray}
 (M^i_j)=\begin{pmatrix}
 \frac{\pi^0}{\sqrt{2}}+\frac{\eta_8}{\sqrt{6}}  &\pi^+ & K^+\\
 \pi^-&-\frac{\pi^0}{\sqrt{2}}+\frac{\eta_8}{\sqrt{6}}&{K^0}\\
 K^-&\overline K^0 &-2\frac{\eta_8}{\sqrt{6}} , 
 \end{pmatrix}+\frac{1}{\sqrt3} \begin{pmatrix}
 \eta_1  &0&0\\
 0& \eta_1  &0\\
 0& 0& \eta_1 , 
 \end{pmatrix},
\end{eqnarray}
The first term forms an SU(3) octet and the second term is a singlet. Grouping  them together it is a nonet of U(3). It is similar for other light mesons, like the vector or axial-vector mesons. 

\subsection{Irreducible Representation Amplitudes}

To obtain irreducible representation amplitudes for $B \to PP$ ($P$ is an element in $M^i_j$) decays, one takes the various representations in Eq.~\eqref{eq:H3615_bb} and uses one $B_i$ and light meson $M^i_j$ to contract all indices in the following manner
\begin{eqnarray}\label{nonisodecomp}
 {\cal A}^{IRA}_t &=&A_3^T B_i (H_{\bar 3})^i (M)_k^j(M)_j^k +C_3^T B_i (M)^i_j (M)^j_k (H_{\bar 3})^k+B_3^T   B_i (H_3)^i (M)_k^k(M)_j^j +D_3^T B_i (M)^i_j  (H_{\bar 3})^j (M)^k_k\nonumber\\
  &&\;\;\;     
  +A_6^T B_i (H_{ 6})^{ij}_k (M)_j^l(M)_l^k 
  +C_6^T B_i (M)^i_j (H_{ 6})^{jl}_k (M)_l^k +B_6^T B_i (H_{ 6})^{ij}_k (M)_j^k(M)_l^l \nonumber\\
  && \;\;\;
  +A_{15}^T B_i (H_{\overline{15}})^{ij}_k (M)_j^l(M)_l^k 
  +C_{15}^T B_i (M)^i_j (H_{\overline{15}})^{jk}_l (M)_k^l +B_{15}^T B_i (H_{\overline{15}})^{ij}_k (M)_j^k(M)_l^l. 
\end{eqnarray}
There also exist  the penguin amplitudes $A^{IRA}_p$ which can be obtained by the replacements    $A_i^T\to A_i^P$, $B_i^T\to B_i^P$,  $C_i^T\to C_i^P$    and $D_i^T\to D_i^P$ ($i=3,6,15$). 

Expanding the above ${\cal A}_t^{IRA}$, one obtains $B\to PP$ amplitudes in the first two columns in Tables \ref{tab:Two_body_Vd} and \ref{tab:Two_body_Vs}. Notice that the amplitude $A_{6}^T$ can be absorbed into $B_{6}^T$ and $C_6^T$ with the following redefinition:
\begin{eqnarray}
C_{6}^{T\prime}= C_{6}^T-A_{6}^T, \;\; B_{6}^{T\prime}= B_{6}^T+A_{6}^T\;.
\end{eqnarray}
Thus we have 18 (tree and penguin contribute 9 each) SU(3) independent  complex amplitudes. Since the phase of one amplitude can be freely chosen,  there are 35 independent parameters to describe the two-body $B\to PP$ decays.  If one also considers $\eta-\eta'$ (or $\eta_8 - \eta_1$) mixing, one more parameter, the mixing angle $\theta$,  is requested making total 36 independent parameters.

\subsection{Topological Diagram Amplitudes}

\begin{figure}
\includegraphics[width=0.7\columnwidth]{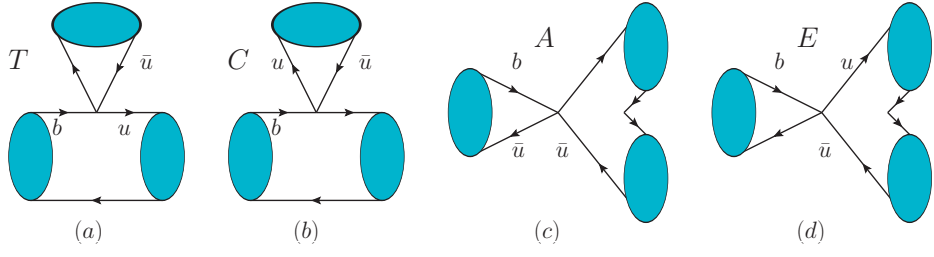}
\caption{Topological  diagrams induced by tree {  amplitudes}. The four panels  denote: the  color-allowed tree amplitude (T), color-suppressed tree amplitude (C), annihilation (A), and W-exchange (E).  }
\label{fig:Feynman_standard}
\end{figure}

The topological diagram amplitudes are obtained by diagrams which connect initial and final states by quark lines as shown in Fig.\ref{fig:Feynman_standard} with vertices determined by the operators in Eq.\eqref{operators}. As shown in many references for instance  Ref.~\cite{Cheng:2014rfa}, they are classified as follows: 
\begin{itemize}
\item[(i)] $T$ denoting the color-allowed tree amplitude with $W$ emission;
\item [(ii)] $C$, denoting the color-suppressed tree diagram;
\item [(iii)] $E$ denoting the $W$-exchange diagram;
\item [(iv)] $P$, corresponding to the QCD penguin contributions;
\item[(v)] $S$, being the flavor singlet QCD penguin;
\item[(vi)] $P_{\rm EW}$, the electroweak penguin. 
\end{itemize}
In addition, there exists annihilation diagrams, usually abbreviated as $A$. In Fig.~\ref{fig:Feynman_standard}, we have only shown the diagrams for tree operators, and those for penguin operators can be derived similarly.  

The electroweak penguins  contain the color-favored contribution $P_{\rm EW}$ and the color-suppressed one $P_{\rm EW}^C$. The electroweak penguin operators can be re-expressed as: 
\begin{eqnarray}
\bar q b \sum_{q'} e_{q'}\bar q'q' =  \bar q b \bar uu - \frac{1}{3} \bar q b \sum_{q'} \bar q'q', \label{relation}
\end{eqnarray} 
where the second part can be incorporated into the penguins transforming as a $\bar 3$ of SU(3).  The contribution from $\bar q b \bar uu $ is similar to tree operators, and thus we will use the symbol $P_{T}$ and $P_{C}$ to denote this electro-weak penguin contribution. The $\bar q b \sum_{q'} \bar q'q'$ is a flavor triplet whose contribution $P'$, as far as flavor SU(3) structure is concerned, can be absorbed into penguin contribution. We can write
\begin{eqnarray}
P_{EW} = P_T -{1\over 3} P'\;,\;\;P^C_{EW} = P_C -{1\over 3}P^{'C}\;.
\end{eqnarray}
The three penguin type of amplitudes  $P$, $P'$ and $P^{'C}$, can be grouped together. We can redefine $P$ by $P+P'+P^{'C}$.

Actually these TDAs can be derived  in a similar way as done for IRAs earlier by indicating $\bar q u \bar u b$ (omitting the Lorentz indices ) by 
$\bar H^{ij}_k $.
For $\Delta S=0$, the non-zero elements are $\bar H^{12}_1= 1$ and for $\Delta S=-1$, $\bar H^{13}_1=1$.  The penguin contribution (including $P$, $P'$ and $P^{'C}$)  is an SU(3) triplet $\bar H^i$ with $\bar H^2 =1$ for the $b\to d$ transition and $H^3$ for the $b\to s$ transition. Eq.~\eqref{relation} implies that the loop induced term proportional to $V^*_{tq}V_{tb}$ has both $\bar H^{ij}_k$ and $\bar H^i$.
{  Note that $\bar H^{ij}_k$ is no longer traceless.}

The tree amplitude is given as 
\begin{eqnarray}
{\cal A}^{TDA}_{t} &=&  T\times  B_i (M)^{i}_j   \bar H^{jl}_k  (M)^k_l   +C\times  B_i (M)^{i}_j \bar H^{lj}_k  (M)^k_l + A \times B_i \bar H^{il}_j   (M)^j_k (M)^{k}_l  + E\times  B_i  \bar H^{li}_j (M)^j_k (M)^{k}_l,\label{eq:tree}
\end{eqnarray}
while 
the penguin amplitude is given  as:
\begin{eqnarray}
{\cal A}^{TDA}_{p}&=& P\times  B_i (M)^{i}_{j} (M)^{j}_k \bar H^k + S\times  B_i (M)^{i}_{j}  \bar H^j (M)^{k}_k+ P_A\times  B_i \bar H^i  (M)^{j}_{k} (M)^{k}_j \nonumber\\
&& + P_{T}\times  B_i (M)^{i}_j   \bar H^{jl}_k  (M)^k_l   +P_C\times  B_i (M)^{i}_j \bar H^{lj}_k  (M)^k_l. 
\end{eqnarray} 
Expanding the above equations,  we obtain the decay amplitudes for $B \to PP$ in the third column in Tables \ref{tab:Two_body_Vd} and \ref{tab:Two_body_Vs}. 
It is necessary to point out that  the singlet contribution in the form $M^j_j$
requires multi-gluon exchanges. One might naively think that its contributions are small compared with other contributions because more gluons are exchanged. However, at energy scale of $B$ decays, the strong couplings are not necessarily very small resulting in non-negligible contributions. One should include them for a complete analysis.



\section{Mismatch and Equivalence }
\label{sec:solution}

From previous discussions, one can see that the total decay amplitudes for $B\to PP$ decays for IRA and TDA can be written as
\begin{eqnarray}
&&{\cal A}^{IRA} = V_{ub}V_{uq}^* {\cal A}^{IRA}_t  + V_{tb}V^*_{tq} {\cal A}^{IRA}_p\;,\nonumber\\
&&{\cal A}^{TDA} = V_{ub}V_{uq}^* {\cal A}^{TDA}_t  + V_{tb}V^*_{tq} {\cal A}^{TDA}_p\;.
\end{eqnarray}

For the amplitudes given in the previous section, it is clear that  for both $A^i_t$ and $A^i_p$, the amplitudes do not have the same number of independent parameters: there are 18 independent complex amplitudes in the IRA, while only 9 amplitudes are included in the TDA. There seems to be a mismatch between the  IRA and TDA approaches. However since both approaches are rooted  in the same basis,  the same physical results should be obtained. It is anticipated that some amplitudes   have been missed and must be added.

\begin{figure}
\includegraphics[width=0.65\columnwidth]{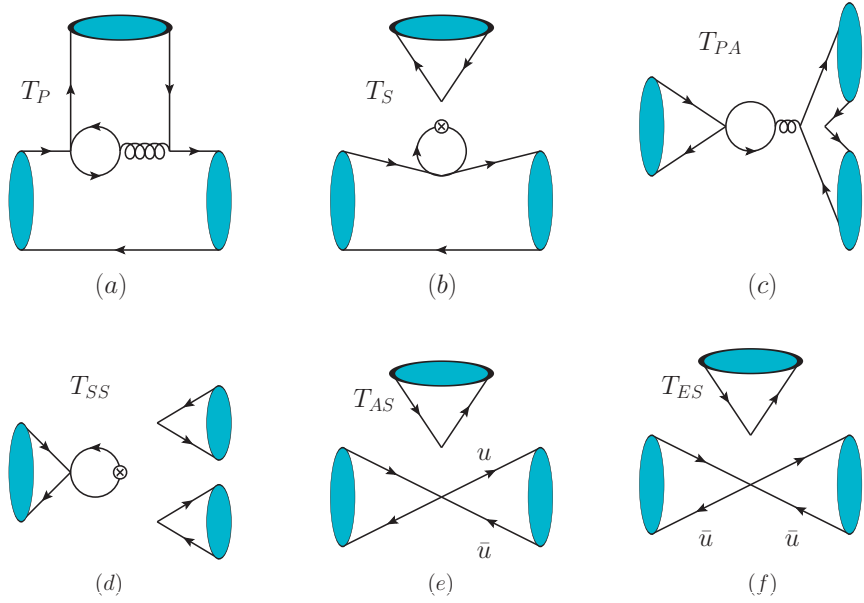}
\caption{Typical  diagrams for the newly introduced amplitudes in Eq.~\eqref{eq:new_tree}.  The crossed vertex denotes the $\bar uu$ annihilation    and the  creation  of two or more gluons.  }
\label{fig:Feynman_new_tree}
\end{figure}

   \begin{table} 
\caption{Decay amplitudes for two-body $B$ decays induced by the $b\to d$ transition.  }\label{tab:Two_body_Vd}\begin{tabular}{cccccc}\hline\hline
{channel} &{IRA}  & {TDA}   \\\hline
$B^-\to \pi^0   \pi^-  $ & $ 4 \sqrt{2} C_{15}^T $ & $ \frac{1}{\sqrt{2}} ({C+T})$  \\\hline
$B^-\to \pi^-   \eta_8  $ & $ \sqrt{\frac{2}{3}} \left(A_6^T+3 A_{15}^T+C_3^T-C_6^T+3 C_{15}^T\right) $ & $\frac{1}{\sqrt{6}}({2 A+C+2 T_P+T} )$  \\\hline
$B^-\to \pi^-   \eta_1  $ & $ \frac{1}{\sqrt{3}} ({2 A_6^T+6 A_{15}^T+3 B_6^T+9 B_{15}^T+2 C_3^T+C_6^T+3 C_{15}^T+3 D_3^T}) $ & $ \frac{1}{\sqrt{3}}({2 A+C+3 T_{\text{ES}}+2 T_P+3 T_S+T}) $ \\\hline
$B^-\to K^0   K^-  $ & $ A_6^T+3 A_{15}^T+C_3^T-C_6^T-C_{15}^T $ & $ A+T_P $  \\\hline
$\overline B^0\to \pi^+   \pi^-  $ & $ 2 A_3^T-A_6^T+A_{15}^T+C_3^T+C_6^T+3 C_{15}^T $ & $ E+T_P+2 T_{PA}+T $  \\\hline
$\overline B^0\to \pi^0   \pi^0  $ & $ 2 A_3^T-A_6^T+A_{15}^T+C_3^T+C_6^T-5 C_{15}^T $ & $ -C+E+T_P+2 T_{PA} $  \\\hline
$\overline B^0\to \pi^0   \eta_8  $ & $ \frac{1}{\sqrt{3}}({-A_6^T+5 A_{15}^T-C_3^T+C_6^T+C_{15}^T}) $ & $ \frac{1}{\sqrt{3}} ({E-T_P})$  \\\hline
$\overline B^0\to \pi^0   \eta_1  $ & $ -\frac{1}{\sqrt{6}}({2 A_6^T-10 A_{15}^T+3 B_6^T-15 B_{15}^T+2 C_3^T+C_6^T-5 C_{15}^T+3 D_3^T}) $ & $ \frac{1}{\sqrt{6}}({3 T_{AS}+2 E-2 T_P-3 T_S} )$  \\\hline
$\overline B^0\to K^+   K^-  $ & $ 2 \left(A_3^T+A_{15}^T\right) $ & $ E+2 T_{PA} $ \\\hline
$\overline B^0\to K^0   \overline K^0  $ & $ 2 A_3^T+A_6^T-3 A_{15}^T+C_3^T-C_6^T-C_{15}^T $ & $ T_P+2 T_{PA} $  \\\hline
$\overline B^0\to \eta_8   \eta_8  $ & $ 2 A_3^T+A_6^T-A_{15}^T+\frac{C_3^T}{3}-C_6^T+C_{15}^T $ & $ \frac{1}{3} \left(C+E+T_P+6 T_{PA}\right) $  \\\hline
$\overline B^0\to \eta_8   \eta_1  $ & $ \frac{1}{3 \sqrt{2}}({-6 A_6^T+6 A_{15}^T-9 B_6^T+9 B_{15}^T+2 C_3^T-3 C_6^T+3 C_{15}^T+3 D_3^T}) $ & $ \frac{1}{3 \sqrt{2}}({3 T_{AS}+2 C+2 E+2 T_P+3 T_S}) $  \\\hline
$\overline B^0\to \eta_1   \eta_1  $ & $ \frac{2}{3} \left(3 A_3^T+9 B_3^T+C_3^T+3 D_3^T\right) $ & $ \frac{2}{3} \left(3 T_{AS}+C+E+T_P+3 T_{PA}+3 T_S+9 T_{SS}\right) $ \\\hline
$\overline B^0_s\to \pi^0   K^0  $ & $ \frac{1}{\sqrt{2}} ({A_6^T+A_{15}^T-C_3^T-C_6^T+5 C_{15}^T})$ & $ \frac{1}{\sqrt{2}}({C-T_P}) $  \\\hline
$\overline B^0_s\to \pi^-   K^+  $ & $ -A_6^T-A_{15}^T+C_3^T+C_6^T+3 C_{15}^T $ & $ T_P+T $  \\\hline
$\overline B^0_s\to K^0   \eta_8  $ & $ \frac{1}{\sqrt{6}}({A_6^T+A_{15}^T-C_3^T-C_6^T+5 C_{15}^T} )$ & $ \frac{1}{\sqrt{6}}({C-T_P}) $  \\\hline
$\overline B^0_s\to K^0   \eta_1  $ & $ -\frac{1}{\sqrt{3}}({2 A_6^T+2 A_{15}^T+3 B_6^T+3 B_{15}^T-2 C_3^T+C_6^T+C_{15}^T-3 D_3^T}) $ & $ \frac{1}{\sqrt{3}}({C+2 T_P+3 T_S}) $  \\\hline
\hline
\end{tabular}
\end{table}

\begin{table}  
\caption{Decay amplitudes for two-body $B$ decays induced by the $b\to s$ transition. }\label{tab:Two_body_Vs}\begin{tabular}{cccccc}\hline\hline
{channel} &{IRA}  & {TDA}   \\\hline
$B^-\to \pi^0   K^-  $ & $ \frac{1}{\sqrt{2}}({A_6^T+3 A_{15}^T+C_3^T-C_6^T+7 C_{15}^T} )$ & $ \frac{1}{\sqrt{2}}({A+C+T_P+T} )$  \\\hline
$B^-\to \pi^-   \overline K^0  $ & $ A_6^T+3 A_{15}^T+C_3^T-C_6^T-C_{15}^T $ & $ A+T_P $ \\\hline
$B^-\to K^-   \eta_8  $ & $ -\frac{1}{\sqrt{6}}({A_6^T+3 A_{15}^T+C_3^T-C_6^T-9 C_{15}^T})$ & $ \frac{1}{\sqrt{6}} ({-A+C-T_P+T})$  \\\hline
$B^-\to K^-   \eta_1  $ & $ \frac{1}{\sqrt{3}}({2 A_6^T+6 A_{15}^T+3 B_6^T+9 B_{15}^T+2 C_3^T+C_6^T+3 C_{15}^T+3 D_3^T}) $ & $ \frac{1}{\sqrt{3}}({2 A+C+3 T_{ {ES}}+2 T_P+3 T_S+T}) $  \\\hline
$\overline B^0\to \pi^+   K^-  $ & $ -A_6^T-A_{15}^T+C_3^T+C_6^T+3 C_{15}^T $ & $ T_P+T $  \\\hline
$\overline B^0\to \pi^0   \overline K^0  $ & $ \frac{1}{\sqrt{2}}({A_6^T+A_{15}^T-C_3^T-C_6^T+5 C_{15}^T}) $ & $ \frac{1}{\sqrt{2}}({C-T_P}) $  \\\hline
$\overline B^0\to \overline K^0   \eta_8  $ & $ \frac{1}{\sqrt{6}} ({A_6^T+A_{15}^T-C_3^T-C_6^T+5 C_{15}^T})$ & $ \frac{1}{\sqrt{6}}({C-T_P}) $  \\\hline
$\overline B^0\to \overline K^0   \eta_1  $ & $ -\frac{1}{\sqrt{3}}({2 A_6^T+2 A_{15}^T+3 B_6^T+3 B_{15}^T-2 C_3^T+C_6^T+C_{15}^T-3 D_3^T}) $ & $ \frac{1}{\sqrt{3}}({C+2 T_P+3 T_S}) $  \\\hline
$\overline B^0_s\to \pi^+   \pi^-  $ & $ 2 \left(A_3^T+A_{15}^T\right) $ & $ E+2 T_{PA} $  \\\hline
$\overline B^0_s\to \pi^0   \pi^0  $ & $ 2 \left(A_3^T+A_{15}^T\right) $ & $ E+2 T_{PA} $  \\\hline
$\overline B^0_s\to \pi^0   \eta_1  $ & $ -\sqrt{\frac{2}{3}} \left(2 A_6^T-4 A_{15}^T+3 B_6^T-6 B_{15}^T+C_6^T-2 C_{15}^T\right) $ & $ \frac{1}{\sqrt{6}} ({3 T_{AS}+C+2 E})$  \\\hline
$\overline B^0_s\to K^+   K^-  $ & $ 2 A_3^T-A_6^T+A_{15}^T+C_3^T+C_6^T+3 C_{15}^T $ & $ E+T_P+2 T_{PA}+T $  \\\hline
$\overline B^0_s\to K^0   \overline K^0  $ & $ 2 A_3^T+A_6^T-3 A_{15}^T+C_3^T-C_6^T-C_{15}^T $ & $ T_P+2 T_{PA} $  \\\hline
$\overline B^0_s\to \eta_8   \eta_8  $ & $ 2 A_3^T-2 A_{15}^T+\frac{4 C_3^T}{3}-4 C_{15}^T $ & $ \frac{1}{3} \left(-2 C+E+4 T_P+6 T_{PA}\right) $ \\\hline
$\overline B^0_s\to \eta_8   \eta_1  $ & $ \frac{1}{3} \sqrt{2} \left(6 A_{15}^T+9 B_{15}^T-2 C_3^T+3 C_{15}^T-3 D_3^T\right) $ & $ -\frac{1}{3 \sqrt{2}}({-3 T_{AS}+C-2 E+4 T_P+6 T_S}) $  \\\hline
$\overline B^0_s\to \eta_1   \eta_1  $ & $ \frac{2}{3} \left(3 A_3^T+9 B_3^T+C_3^T+3 D_3^T\right) $ & $ \frac{2}{3} \left(3 T_{AS}+C+E+T_P+3 T_{PA}+3 T_S+9 T_{SS}\right) $  \\\hline
\hline
\end{tabular}
\end{table}

A close inspection shows that several topological diagrams were not included in the previous TDA analysis. For the tree amplitudes we show  the relevant diagrams in Fig.~\ref{fig:Feynman_new_tree}.  The missing  penguin diagrams can be obtained similarly. Since there are electroweak penguin operator contributions, as far as the SU(3) irreducible components are concerned, the effective Hamiltonian have the same SU(3) structure as the tree contributions.
Taking  these contributions into account, we have  the following topological amplitudes: 
\begin{eqnarray}
{\cal A}_{t}^{' TDA}&=&  T_{S} B_i  (M)^{i}_j  \bar H^{lj}_{l}  (M)^k_k +T_{P} B_i (M)^{i}_j   (M)^j_k \bar H^{lk}_{l} + T_{PA} B_i \bar H^{li}_{l}  (M)^j_k (M)^{k}_j   + T_{SS} B_i \bar H^{li}_{l}  (M)^j_j (M)^{k}_{k}   \nonumber\\
&& +T_{AS} B_i \bar H^{ji}_{l}  (M)^{l}_j  (M)^k_k+T_{ES} B_i  \bar H^{ij}_{l}   (M)^{l}_j    (M)^k_k,  \label{eq:new_tree}\\
{\cal A}_{p}^{'TDA}&=&  P_{SS} B_i \bar H^i  (M)^{j}_{j} (M)^{k}_k+ P_{TA} B_i \bar H^{il}_j   (M)^j_k (M)^{k}_l  + P_{TE} B_i  \bar H^{ji}_k   (M)^k_l (M)^{l}_j \nonumber\\
&& +P_{AS} B_i \bar H^{ji}_{l}  (M)^{l}_j  (M)^k_k+P_{ES} B_i  \bar H^{ij}_{l}   (M)^{l}_j    (M)^k_k. \label{eq:new_penguin}
\end{eqnarray}
The   mismatch problem can be partly traced to the fact that $\bar H^{ij}_k$ defined in the TDA analysis is not traceless, that is $\bar H^{lj}_l \neq 0$. Because of this fact,  $B_i$ and the two $M^i_j$ can contract with $\bar H^{lj}_l$ to
form SU(3) invariant amplitudes and also the trace for $M^i_j$ is not zero when $\eta_1$ is included in the final states. While in the previous discussions, these terms are missed. 

One can expand the above new terms to obtain the results for   tree amplitudes  in Tables \ref{tab:Two_body_Vd} and \ref{tab:Two_body_Vs}.   With these new amplitudes at hand, one can derive the relation  between the two sets of amplitudes: 
\begin{eqnarray}
&&A_3^T= -\frac{A}{8} + \frac{3E}{8}+T_{PA}, \;\hspace{2.3cm}
B_3^T=  T_{SS} +\frac{3T_{AS}-T_{ES}}{8},\;\;\;\;\;\;\;\;\;
C_3^T=  \frac{1}{8} ({3A-C-E+3T})+T_P,  \nonumber\\
&&D_3^T=  T_{S} +\frac{1}{8} (3C-T_{AS}+3T_{ES}-T),\;\; B_6^{\prime T}=  \frac{1}{4}(A-E+T_{ES}-T_{AS}), \;\; C_6^{\prime T}=  \frac{1}{4}(-A-C+E+T),\nonumber\\
&&A_{15}^T=  \frac{A+E}{8}, \hspace{4cm}
B_{15}^T= \frac{T_{ES}+T_{AS}}{8}, \hspace{2cm} 
C_{15}^T=  \frac{C+T}{8}. \label{eq:relation_TDA2IRA}
\end{eqnarray} 
Here we have absorbed  the $A_{6}^T$ into $B_{6}^{\prime T}$ and $C_{6}^{\prime T}$. In the appendix, we give a direct derivation of relations between IRA and TDA amplitudes, in which the amplitude  $A_{6}^T$ is kept. 

Naively there are total 10 tree amplitudes and 10 penguin   amplitudes defined in Eq.~(\ref{eq:tree},\ref{eq:new_tree}). However, only 9 of the 10 tree amplitudes are independent. Choosing the option to eliminate the W-exchange $E$, we can express the TDA amplitudes in terms of the IRA ones:   
\begin{eqnarray}
&&T+E = 4A_{15}^T +2C_{6}^{\prime T} +4C_{15}^T, \hspace{1.3cm} C-E=-4A_{15}^T -2C_{6}^{\prime T} +4C_{15}^T, \nonumber\\
&&A+E = 8A_{15}^T , \hspace{3.7cm} T_{P}-E= -5A_{15}^T +C_{3}^T-C_{6}^{\prime T} -C_{15}^T, \nonumber\\
&&T_{PA}+ \frac{E}{2}= A_{3}^T+A_{15}^T , \hspace{2.5cm} T_{AS}+E = 4A_{15}^T -2B_{6}^{\prime T} +4B_{15}^T, \nonumber\\
&&T_{ES}-E = -4A_{15}^T +2B_{6}^{\prime T} +4B_{15}^T, \;\; \;\;\;\;\;T_{SS}- \frac{E}{2} = -2A_{15}^T +B_{3}^T +B_{6}^{\prime T} -B_{15}^T, \nonumber\\
&&T_{S}+E = 4A_{15}^T -B_{6}^{\prime T} -B_{15}^T +C_{6}^{\prime T} -C_{15}^T +D_{3}^T. 
\end{eqnarray}
 The analysis  of  penguin contributions is similar with the replacement for TDA amplitudes: 
 \begin{eqnarray}
 T\to P_{T}, \;\; C\to P_{C}, \;\; A\to P_{TA}, \;\; T_{P} \to P,\;\; E\to P_{TE},  \nonumber\\
 T_{PA}\to P_A, \;\; T_{AS}\to P_{AS}, \;\; T_{ES}\to P_{ES}, \;\; T_{SS}\to P_{SS}, \;\; T_{S}\to S.  
 \end{eqnarray}
 
From the above discussions we see that the two sets of amplitudes in IRA and TDA can be mutually expressed by each other. The IRA and TDA approaches are completely equivalent. As long as all amplitudes are taken into account in the analysis, they give the same results for $B\to PP$ decays, and we expect the equivalence for other decays~\footnote{In a recent  study~\cite{Chua:2018ikx}, TDA amplitudes have been obtained. However, the independence of  amplitudes is not  discussed in TDA approach.  }. 

\section{Discussions and Conclusions}
\label{sec:conclusion}

{  We now make a few remarks about our results obtained.}

Several missing terms in the TDA analysis involve the trace $M^{j}_{j}$. The trace actually singles out the singlet in the nonet representation $M^i_j$. To have a color singlet in the diagram shown in Figs.~\ref{fig:Feynman_standard} and \ref{fig:Feynman_new_tree}, the single $M^j_j$ need to exchange two or more gluons. As pointed out earlier that these contributions are expected to be small compared with other contributions. However, at energy scale of $B$ decays, the strong couplings are not necessarily very small resulting in non-negligible contributions. Terms associated with the trace $\bar H^{lj}_l$ actually can be thought of as turning the tree operator into penguin operator with $u$ quark exchange in the loop whose Wilson coefficient contains the large logarithms  $ln(\mu/m_u)$ which can also make non-negligible contributions.  One should include them for a complete analysis. 

Recently, 
Ref.~\cite{Hsiao:2015iiu} has performed a fit of $B\to PP$ decays in the IRA scheme. Depending on various options to use the data, four cases are considered in Ref.~\cite{Hsiao:2015iiu}. As an example, we quote the results in their  case 4: 
\begin{eqnarray}
|C_{\bar 3}^T|= -0.211\pm 0.027,\;\; \delta_{\bar 3}^T= (-140\pm6)^\circ,  \;\; |B_{\overline {15}}^T| = -0.038\pm 0.016, \;\; \delta_{B_{\overline {15}}^T}= (78\pm 48)^\circ,  \label{ea:IRA_fit}
\end{eqnarray}
where the magnitudes and  strong phases relative to $C_{\bar 3}^P$ have been given. From Eq.~\eqref{eq:relation_TDA2IRA}, one can see that  the $C_{\bar 3}^T$ is a combination of color-allowed tree $T$,  color-suppressed tree amplitude  $C$ and others while  the $B_{\overline{15}}^T$ corresponds to $(T_{ES} + T_{TS)}/8$ in TDA approach.  The fitted result in Eq.~\eqref{ea:IRA_fit} indicates that compared to $C_{\bar 3}^T$, the $B_{\overline {15}}^T$ can reach $20\%$ in magnitude, and more importantly, the strong phases are different significantly.  The $B_{\overline {15}}^T$, equivalently  $T_{ES}$ and $T_{TS}$,  have non-negligible contributions supporting our call for a complete analysis.
With more and more accurate data for $B\to PP$ from experiments, one can now carry out a more careful analysis to obtain the amplitudes and  derive   implications for model calculations of the relevant amplitudes.

Without the new contributions in the TDA analysis, some of the amplitudes only have terms proportional to $V_{tq}^*V_{tb}$, such as $\overline B^0 \to K^0 \bar K^0$ and $\overline B^0_s \to K^0 \bar K^0$. This implies that  CP violation in these two decays are identically zero. However, these two decay modes receive contributions from the new terms $T_P + 2 T_{PA}$ which is multiplied by $V_{uq}^*V_{ub}$. In principle they can have non-zero CP violation. {  Therefore if one takes into account the missing tree and penguin amplitudes,  an important consequence is that no charmless and hadronic  $B$ decay channel   has a vanishing direct CP asymmetry.}

Flavor SU(3) symmetry is an approximate symmetry, and  symmetry breaking sources exist in QCD, mostly caused by the unequal masses for the light $u,d,s$ quarks. {  How SU(3) breaking effect manifest itself is not completely clear. Experimental data ~\cite{Aaij:2013iua,Patrignani:2016xqp} for $\overline B^0\to K^-\pi^+$ and $\overline B_s\to K^+\pi^-$ agree with relation predicted for these two modes under SU(3) symmetry~\cite{He:1998rq, He:2013vta}. Therefore the use of SU(3) symmetry for $B$ decays  might be justified. With new data from BELLE II and LHCb, one can study SU(3) symmetry in $B$ decays with a high precisions. One should keep in mind that
 for an appropriate analysis of SU(3) symmetry breaking, one must take into account all the above amplitudes, otherwise, the missing amplitudes will be disguised  as symmetry breaking effects. As we have shown above, the modification due to the missing amplitudes  can reach $20\%$, which is comparable with  the generic SU(3) symmetry breaking effects.  Thus the additional TDAs must be treated carefully   to correctly interpret the data.}


Our analysis is also applicable to other decay channels of heavy mesons and   baryons.  In the appendix, we give a discussion on  the $D$ meson decays.  For charm quark decay, penguin operators are often negligible and the $\bar 3$ representation does not contribute either.  So there are five independent tree amplitudes, while in TDA only four amplitudes, $T,C,E,A$, are used for the global fit.

In summary, we have carried out an analysis comparing two different approaches, the irreducible representation amplitude and topological diagram amplitude, to study $B \to PP$ decays. We find that previous analyses in the literature using these two methods  do not match consistently in several ways. A few SU(3) independent amplitudes have been overlooked in the TDA approach.  Taking these new amplitudes into account, we find a consistent description in both approaches. These new  amplitudes can affect  direct CP asymmetries in some channels significantly. A consequence is that for any charmless hadronic  decays of heavy mesons, the direct CP symmetry cannot be identically zero.  With more data become available, we can have a a better understanding of the role of flavor SU(3) symmetry in $B$ decays.

\section*{Acknowledgement}

The authors are   grateful to Cheng-Wei Chiang  for useful discussions and valuable comments.
WW thanks the hospitality from   Tsinghua University  at Hsinchu where this work was finalized. 
This work is supported  in part   by National  Natural
Science Foundation of China under Grant
 No.11575110, 11575111, 11655002, 11735010,  Natural  Science Foundation of Shanghai under Grant  No.~15DZ2272100, and by   MOST (Grant No. MOST104-2112-M-002-015-MY3 and 106-2112-M-002-003-MY3). 

\begin{appendix}
\section{A derivation of Decomposition}

Using  $O^{12}_1=\bar u b \bar d u$ as an   example, we have the decomposition of tree operator: 
\begin{eqnarray}
 O^{12}_1 &=& \frac{1}{8} O_{\overline{15}}+ \frac{1}{4} O_{ 6} -\frac{1}{8} O_{\bar3} + \frac{3}{8} O_{\bar3'}, 
\end{eqnarray}
with  
\begin{eqnarray} 
 O_{\overline{15}} &=& 3 \bar u b \bar du + \bar d b \bar uu - 2\bar db \bar dd - \bar s b \bar d s - \bar d b\bar ss, \;\;
 O_{ 6}= \bar u b \bar du - \bar d b \bar ud - \bar s b \bar ds + \bar d b \bar ss, \nonumber\\
 O_{\bar 3} &=& \bar d b \bar uu + \bar db \bar dd + \bar db \bar ss, \;\;
 O_{\bar3'} = \bar u b \bar du + \bar db \bar dd + \bar sb \bar ds. 
\end{eqnarray}
It implies: 
\begin{eqnarray}
\bar  H^{ij}_k = \frac{1}{8} (H_{\overline{15}})^{ij}_k +\frac{1}{4} (H_{  6})^{ij}_k -\frac{1}{8} (H_{\overline{3}})^i \delta^{j}_k + \frac{3}{8}(H_{\overline{3}'})^{j} \delta^i_k. 
\end{eqnarray}

Substituting this expression into the amplitude $T$ for instance, we have  
\begin{eqnarray}
T\times  B_i (M)^{i}_j  \bar H^{jl}_k  (M)^k_l   = T\times  B_i (M)^{i}_j (M)^k_l  \times \left(\frac{1}{8} (H_{15})^{jl}_k +\frac{1}{4} (H_{\bar 6})^{jl}_k -\frac{1}{8} (H_3)^j \delta^{l}_k + \frac{3}{8}(H_{3'})^{l} \delta^j_k\right), 
\end{eqnarray}
contributing to 
\begin{eqnarray}
C_{15}^T&=& \frac{1}{8}T + ..., \;\;
C_{6}^T = \frac{1}{4} T + ...,\;\;
C_{3}^T =\frac{3}{8} T + ...,\;\;
D_{3}^T = -\frac{1}{8} T + ... \;\;\; . 
\end{eqnarray} 
Others TDA amplitudes  can be analyzed  similarly, and thus  
one has  
\begin{eqnarray}
A_3^T= -\frac{A}{8} + \frac{3E}{8}+T_{PA}, && 
B_3^T=  T_{SS} +\frac{3T_{AS}-T_{ES}}{8},\nonumber\\
C_3^T=  \frac{1}{8} ({3A-C-E+3T})+T_P, &&  D_3^T=  T_{S} +\frac{1}{8} (3C-T_{AS}+3T_{ES}-T), \nonumber\\
A_6^T=  \frac{1}{4}(A-E),  &&
B_6^T=  \frac{1}{4}(T_{ES}-T_{AS}),\nonumber\\
C_6^T=  \frac{1}{4}(-C+T),  &&
A_{15}^T=  \frac{A+E}{8},  \nonumber\\
B_{15}^T= \frac{T_{ES}+T_{AS}}{8}, &&
C_{15}^T=  \frac{C+T}{8}. 
\end{eqnarray}
The inverse relation is given as:
\begin{eqnarray}
T&=& 2 C_6^T +4C_{15}^T, \;\; C= 4C_{15}^T -2C_{6}^T, \;\; A= 2A_6^T +4A_{15}^T,\;\; E= 4A_{15}^T-2A_6^T,\nonumber\\
T_{P}&=& -A_{6}^T-A_{15}^T+C_{3}^T-C_{6}^T-C_{15}^T, \;\; T_{PA}= A_{3}^T+A_{6}^T -A_{15}^T,\;\; T_{AS}= 4B_{15}^T -2B_{6}^T, \nonumber\\
T_{ES}&=& 2B_6^T +4B_{15}^T, \;\; T_{SS}=B_{3}^T+B_{6}^T -B_{15}^T,\;\;T_{S}=-B_{6}^T-B_{15}^T+C_{6}^T -C_{15}^T +D_{3}^T. 
\end{eqnarray}
From the expansion of IRA amplitudes, one can notice that the $A_{6}^T$ can be absorbed into $B_{6}^T$ and $C_{6}^T$. 

\section{$D$ meson decays} 

  \begin{table}
\caption{Decay amplitudes for two-body Cabibblo-Allowed $D$ decays.}\label{tab:Two_body_Cabibblo_Allowed}\begin{tabular}{cccccc}\hline\hline
{channel} &{IRA}  & {TDA}   \\\hline
$D^0\to \pi^+   K^-  $ & $ -A_6^T+A_{15}^T+C_6^T+C_{15}^T $ & $ E+T $  \\\hline
$D^0\to \pi^0   \overline K^0  $ & $ \frac{1}{\sqrt{2}}({A_6^T-A_{15}^T-C_6^T+C_{15}^T}) $ & $ \frac{1}{\sqrt{2}} ({C-E})$  \\\hline
$D^0\to \overline K^0   \eta_8  $ & $ \frac{1}{\sqrt{6}}({A_6^T-A_{15}^T-C_6^T+C_{15}^T})$ & $ \frac{1}{\sqrt{6}} ({C-E}$) & \\\hline
$D^0\to \overline K^0   \eta_1  $ & $ \frac{1}{\sqrt{3}}({-2 A_6^T+2 A_{15}^T-3 B_6^T+3 B_{15}^T-C_6^T+C_{15}^T}) $ & $ \frac{1}{\sqrt{3}}({3 T_{{AS}}+C+2 E}) $  \\\hline
$D^+\to \pi^+   \overline K^0  $ & $ 2 C_{15}^T $ & $ C+T $  \\\hline
$D^+_s\to \pi^+   \eta_8  $ & $ \sqrt{\frac{2}{3}} \left(A_6^T+A_{15}^T-C_6^T-C_{15}^T\right) $ & $ \sqrt{\frac{2}{3}} (A-T) $ \\\hline
$D^+_s\to \pi^+   \eta_1  $ & $ \frac{1}{\sqrt{3}}({2 A_6^T+2 A_{15}^T+3 B_6^T+3 B_{15}^T+C_6^T+C_{15}^T}) $ & $ \frac{1}{\sqrt{3}}({2 A+3 T_{ {ES}}+T})$  \\\hline
$D^+_s\to K^+   \overline K^0  $ & $ A_6^T+A_{15}^T-C_6^T+C_{15}^T $ & $ A+C $  \\\hline
\hline
\end{tabular}
\end{table}
\begin{table}
\caption{Decay amplitudes for two-body Singly Cabibblo-Suppressed $D$ decays.}\label{tab:Two_body_Singly_Cabibblo_Suppressed}\begin{tabular}{cccccc}\hline\hline
{channel} &{IRA}  & {TDA}   \\\hline
$D^0\to \pi^+   \pi^-  $ & $ \sin\theta_C \left(A_6^T-A_{15}^T-C_6^T-C_{15}^T\right) $ & $ -\sin\theta_C (E+T) $  \\\hline
$D^0\to \pi^0   \pi^0  $ & $ \sin\theta_C \left(A_6^T-A_{15}^T-C_6^T+C_{15}^T\right) $ & $ \sin\theta_C (C-E) $  \\\hline
$D^0\to \pi^0   \eta_8  $ & $ -\frac{1}{\sqrt{3}}{\sin\theta_C \left(A_6^T-A_{15}^T-C_6^T+C_{15}^T\right)} $ & $ \frac{1}{\sqrt{3}}{\sin\theta_C (E-C)} $  \\\hline
$D^0\to \pi^0   \eta_1  $ & $ -\frac{1}{\sqrt{6}}{\sin\theta_C \left(2 A_6^T-2 A_{15}^T+3 B_6^T-3 B_{15}^T+C_6^T-C_{15}^T\right)} $ & $ \frac{1}{\sqrt{6}}{\sin\theta_C \left(3 T_{{AS}}+C+2 E\right)} $  \\\hline
$D^0\to K^+   K^-  $ & $ \sin\theta_C \left(-A_6^T+A_{15}^T+C_6^T+C_{15}^T\right) $ & $ \sin\theta_C (E+T) $  \\\hline
$D^0\to \eta_8   \eta_8  $ & $ -\sin\theta_C \left(A_6^T-A_{15}^T-C_6^T+C_{15}^T\right) $ & $ \sin\theta_C (E-C) $ \\\hline
$D^0\to \eta_8   \eta_1  $ & $ \frac{1}{\sqrt{2}}{\sin\theta_C \left(2 A_6^T-2 A_{15}^T+3 B_6^T-3 B_{15}^T+C_6^T-C_{15}^T\right)} $ & $ -\frac{1}{\sqrt{2}}{\sin\theta_C \left(3 T_{{AS}}+C+2 E\right)} $  \\\hline
$D^+\to \pi^+   \pi^0  $ & $ \sqrt{2} \sin\theta_C C_{15}^T $ & $ \frac{1}{\sqrt{2}}{\sin\theta_C (C+T)} $ \\\hline
$D^+\to \pi^+   \eta_8  $ & $ -\sqrt{\frac{2}{3}} \sin\theta_C \left(A_6^T+A_{15}^T-C_6^T+2 C_{15}^T\right) $ & $ -\frac{1}{\sqrt{6}}{\sin\theta_C (2 A+3 C+T)} $  \\\hline
$D^+\to \pi^+   \eta_1  $ & $ -\frac{1}{\sqrt{3}}{\sin\theta_C \left(2 A_6^T+2 A_{15}^T+3 B_6^T+3 B_{15}^T+C_6^T+C_{15}^T\right)} $ & $ -\frac{1}{\sqrt{3}}{\sin\theta_C \left(2 A+3 T_{\text{ES}}+T\right)} $  \\\hline
$D^+\to K^+   \overline K^0  $ & $ -\sin\theta_C \left(A_6^T+A_{15}^T-C_6^T-C_{15}^T\right) $ & $ \sin\theta_C (T-A) $  \\\hline
$D^+_s\to \pi^+   K^0  $ & $ \sin\theta_C \left(A_6^T+A_{15}^T-C_6^T-C_{15}^T\right) $ & $ \sin\theta_C (A-T) $  \\\hline
$D^+_s\to \pi^0   K^+  $ & $ \frac{1}{\sqrt{2}} {\sin\theta_C \left(A_6^T+A_{15}^T-C_6^T+C_{15}^T\right)}$ & $ \frac{1}{\sqrt{2}}{\sin\theta_C (A+C)} $  \\\hline
$D^+_s\to K^+   \eta_8  $ & $ -\frac{1}{\sqrt{6}}{\sin\theta_C \left(A_6^T+A_{15}^T-C_6^T+5 C_{15}^T\right)} $ & $ -\frac{1}{\sqrt{6}}{\sin\theta_C (A+3 C+2 T)} $  \\\hline
$D^+_s\to K^+   \eta_1  $ & $ \frac{1}{\sqrt{3}} {\sin\theta_C \left(2 A_6^T+2 A_{15}^T+3 B_6^T+3 B_{15}^T+C_6^T+C_{15}^T\right)}$ & $ \frac{1}{\sqrt{3}}{\sin\theta_C \left(2 A+3 T_{\text{ES}}+T\right)} $ \\\hline
\hline
\end{tabular}
\end{table}
\begin{table}
\caption{Decay amplitudes for two-body Doubly Cabibblo-Suppressed $D$ decays.}\label{tab:Two_body_Doubly_Cabibblo_Suppressed}\begin{tabular}{cccccc}\hline\hline
{channel} &{IRA}  & {TDA}   \\\hline
$D^0\to \pi^0   K^0  $ & $ -\frac{1}{\sqrt{2}}{\sin^2\theta_C \left(A_6^T-A_{15}^T-C_6^T+C_{15}^T\right)} $ & $ -\frac{1}{\sqrt{2}}{\sin^2\theta_C (C-E)} $  \\\hline
$D^0\to \pi^-   K^+  $ & $ -\sin^2\theta_C \left(-A_6^T+A_{15}^T+C_6^T+C_{15}^T\right) $ & $ -\sin^2\theta_C (E+T) $  \\\hline
$D^0\to K^0   \eta_8  $ & $ -\frac{1}{\sqrt{6}}{\sin^2\theta_C \left(A_6^T-A_{15}^T-C_6^T+C_{15}^T\right)} $ & $- \frac{1}{\sqrt{6}}{\sin^2\theta_C (C-E)} $  \\\hline
$D^0\to K^0   \eta_1  $ & $ \frac{1}{\sqrt{3}} {\sin^2\theta_C \left(2 A_6^T-2 A_{15}^T+3 B_6^T-3 B_{15}^T+C_6^T-C_{15}^T\right)}$ & $ -\frac{1}{\sqrt{3}} {\sin^2\theta_C \left(3 T_{{AS}}+C+2 E\right)}$  \\\hline
$D^+\to \pi^+   K^0  $ & $ -\sin^2\theta_C \left(A_6^T+A_{15}^T-C_6^T+C_{15}^T\right) $ & $ -\sin^2\theta_C (A+C) $  \\\hline
$D^+\to \pi^0   K^+  $ & $ -\frac{1}{\sqrt{2}}{\sin^2\theta_C \left(A_6^T+A_{15}^T-C_6^T-C_{15}^T\right)} $ & $ -\frac{1}{\sqrt{2}} {\sin^2\theta_C (A-T)}$  \\\hline
$D^+\to K^+   \eta_8  $ & $\frac{1}{\sqrt{6}}{\sin^2\theta_C \left(A_6^T+A_{15}^T-C_6^T-C_{15}^T\right)} $ & $ -\frac{1}{\sqrt{6}} {\sin^2\theta_C (T-A)}$ \\\hline
$D^+\to K^+   \eta_1  $ & $- \frac{1}{\sqrt{3}}{\sin^2\theta_C \left(2 A_6^T+2 A_{15}^T+3 B_6^T+3 B_{15}^T+C_6^T+C_{15}^T\right)} $ & $- \frac{1}{\sqrt{3}}{\sin^2\theta_C \left(2 A+3 T_{\text{ES}}+T\right)} $  \\\hline
$D^+_s\to K^+   K^0  $ & $- 2 \sin^2\theta_C C_{15}^T $ & $- \sin^2\theta_C (C+T) $  \\\hline
\hline
\end{tabular}
\end{table}

  The effective Hamiltonian for charm quark decay is given as 
\begin{eqnarray}
{\cal H}_{eff} &=& \frac{G_F}{\sqrt2}\big\{V_{cs} V_{ud}^* [C_1O_1^{sd}+C_2O_2^{sd}]+V_{cd} V_{ud}^* [C_1O_1^{dd}+C_2O_2^{dd}] \nonumber\\
&&+ V_{cs} V_{us}^* [C_1O_1^{ss}+C_2O_2^{ss}]+V_{cd} V_{us}^* [C_1O_1^{ds}+C_2O_2^{ds}]  \big\}, 
\end{eqnarray}
where 
\begin{eqnarray}
O_1^{sd} &=& [\bar s^i  \gamma_{\mu}(1-\gamma_5) c^j ][\bar u^i \gamma^{\mu}(1-\gamma_5) d^j], \;\; 
O_2^{sd} = [\bar s  \gamma_{\mu}(1-\gamma_5) c][\bar u \gamma^{\mu}(1-\gamma_5) d], 
\end{eqnarray}
and other operators can be obtained by replacing the $d,s$ quark fields. 
In the above equations, we have neglected the   highly-suppressed
penguin contributions.
Tree operators  transform
under the flavor SU(3) symmetry as ${\bf \bar  3}\otimes {\bf  3}\otimes {\bf  \bar
3}={\bf   \bar 3}\oplus {\bf   \bar3}\oplus {\bf  6}\oplus {\bf   {\overline {15}}}$. 
For the $c\to s  u \bar d$ transition, we have
\begin{eqnarray}
(H_{  6})^{31}_2=-(H_{  6})^{13}_2=1,\;\;\;
 (H_{\overline {15}})^{31}_2= (H_{\overline {15}})^{13}_2=1,\label{eq:H3615_c_allowed}
\end{eqnarray}
while for the  doubly Cabibbo suppressed $c\to d  u \bar s$ transition, we have
\begin{eqnarray}
(H_{  6})^{21}_3=-(H_{  6})^{12}_3=-\sin^2\theta_C,\;\;
 (H_{\overline {15}})^{21}_3= (H_{\overline {15}})^{12}_3=-\sin^2\theta_C. \label{eq:H3615_c_doubly_suprressed}
\end{eqnarray}
{  The CKM matrix elements for  $c\to u \bar dd$ and $c\to u \bar ss$ transitions are approximately equal in magnitude but different in sign: $V_{cd}V_{ud}^* = - V_{cs}V_{us}^* -V_{cb}V_{ub}^* \approx - V_{cs}V_{us}^*$ (accurate at   $10^{-3}$). With both contributions,  the  contributions from the $\bar 3$ representation vanish, and   one has the nonzero components:}
\begin{eqnarray}
(H_{  6})^{31}_3 =-(H_{  6})^{13}_3 =(H_{  6})^{12}_2 =-(H_{  6})^{21}_2 =\sin(\theta_C),\nonumber\\
 (H_{\overline {15}})^{31}_3= (H_{\overline  {15}})^{13}_3=-(H_{\overline  {15}})^{12}_2=-(H_{\overline  {15}})^{21}_2= \sin(\theta_C).\label{eq:H3615_cc_singly_suppressed}
\end{eqnarray}

A few remarks are in order. 
\begin{itemize} 

\item The expanded amplitudes are given in Tab.~\ref{tab:Two_body_Cabibblo_Allowed} for Cabibbo-allowed channels, Tab.~\ref{tab:Two_body_Singly_Cabibblo_Suppressed} for singly Cabibbo-suppressed modes, and Tab.~\ref{tab:Two_body_Doubly_Cabibblo_Suppressed} for doubly  Cabibbo-suppressed  decay channels.

\item  

One can derive the following relations between the two sets of amplitudes:  
\begin{eqnarray}
A_{6}^T&=& \frac{1}{2} (A-E), \;\; 
A_{15}^T= \frac{1}{2} (A+E), \;\;
B_{6}^T= \frac{1}{2} (T_{ES}-T_{AS}), \nonumber\\ 
B_{15}^T&=& \frac{1}{2}(T_{ES}+T_{AS}), \;\;
C_{6}^T= \frac{1}{2} (T-C), \;\; 
C_{15}^T= \frac{1}{2}(T+C). 
\end{eqnarray}

\item The amplitudes $A_{6}^T$ can be incorporated in  $B_{6}^{T\prime}$ and $C_{6}^{T\prime}$, and then we have 
\begin{eqnarray} 
A_{15}^T&=& \frac{1}{2} (A+E), \nonumber\\
B_{6}^{\prime T}&=& \frac{1}{2} (T_{ES}-T_{AS}+A-E), \;\; 
B_{15}^T= \frac{1}{2}(T_{ES}+T_{AS}), \nonumber\\
C_{6}^{\prime T}&=& \frac{1}{2} (T-C-A+E), \;\; 
C_{15}^T= \frac{1}{2}(T+C),  
\end{eqnarray}
with the inverse relation:
\begin{eqnarray}
 T &=& A_{15}^T + C_{6}^{\prime T}+C_{15}^T-E, \;\;  C= -A_{15}^T - C_{6}^{\prime T}+C_{15}^T+ E, \;\;
 A= 2A_{15}^T-E, \nonumber\\
 T_{ES}&=& -A_{15}^T +B_{6}^{\prime T} +B_{15}^T +E, \;\;  T_{AS}= A_{15}^T -B_{6}^{\prime T} +B_{15}^T -E. 
\end{eqnarray}

\end{itemize}

\end{appendix}

\end{document}